\documentclass[journal,twoside,web]{IEEE/ieeecolor}

\usepackage{amsmath,amssymb,amsfonts}
\usepackage{amsthm}
\usepackage{IEEE/generic}
\usepackage{cite}
\usepackage{graphicx}
\usepackage{algorithm,algorithmic}
\usepackage{hyperref}
\hypersetup{hidelinks=true}
\usepackage{textcomp}
\usepackage{todonotes}
\usepackage{siunitx}
\usepackage{physics}
\usepackage[switch,columnwise]{lineno}

\def\BibTeX{{\rm B\kern-.05em{\sc i\kern-.025em b}\kern-.08em
    T\kern-.1667em\lower.7ex\hbox{E}\kern-.125emX}}

\theoremstyle{definition}
\newtheorem{definition}{Definition}

\theoremstyle{plain}

\theoremstyle{plain}
\newtheorem{fact}{Fact}

\begin{document}

\title{Ankle Exoskeletons May Hinder Standing Balance in Simple Models of Older and Younger Adults}
\author{Daphna Raz, Varun Joshi, Brian R. Umberger, and Necmiye Ozay \IEEEmembership{Senior Member, IEEE}
\thanks{This work was supported by NIH grants F31-EB032745 and R01AG068102 and NSF grant CPS-1931982} %
\thanks{ D. Raz and N. Ozay are with the University of Michigan Robotics Department, Ann Arbor, MI USA (e-mail: daphraz@umich.edu, necmiye@umich.edu). }
\thanks{V. Joshi and B. R. Umberger are with the University of Michigan School of Kinesiology, Ann Arbor, MI USA (e-mail: varunjos@umich.edu, umberger@umich.edu).}}

\maketitle

\begin{abstract}
Humans rely on ankle torque to maintain standing balance, particularly in the presence of small to moderate perturbations. Reductions in maximum torque (MT) production and maximum rate of torque development (MRTD) occur at the ankle with age, diminishing stability. Ankle exoskeletons are powered orthotic devices that may assist older adults by compensating for reduced torque and power production capabilities. They may also be able to assist with ankle strategies used for balance. However, the effect of such devices on standing balance in older adults is not well understood. Here, we model the effects ankle exoskeletons have on stability in physics-based models of healthy young and old adults, focusing on the potential to mitigate age-related deficits in MT and MRTD. Using backward reachability, a mathematical technique for analyzing the behavior of dynamical systems, we compute the set of stable center of mass positions and velocities for sex and age adjusted models of human standing balance with an ankle exoskeleton. We show that an ankle exoskeleton moderately reduces feasible stability boundaries in users who have full ankle strength. For individuals with age-related deficits, there is a trade-off. While exoskeletons augment stability at low center of mass velocities, they reduce stability in some high velocity conditions. Our results suggest that well-established control strategies must still be experimentally validated in older adults.
\end{abstract}

\begin{IEEEkeywords}
Aging, biomechanics, exoskeletons, human stability, standing balance.
\end{IEEEkeywords}

\section{Introduction}
\label{sec:introduction}
Ankle exoskeletons are powered orthotic devices that can assist people with mobility impairments, such as older adults, by compensating for reduced muscle torque and power capabilities. Typically, they provide actuation about the ankle joint only in the sagittal plane, generating a push-off torque during dynamic activities such as walking and stair climbing. Exoskeletons have been shown to reduce metabolic cost of transport during walking in both younger \cite{Poggensee2021} and older adults \cite{Lakmaa2024}. The main appeal of lower-limb exoskeletons to older adults, however, is their perceived potential to reduce fall risk  \cite{Raitor2024}. It is therefore crucial to develop our understanding of the effects of ankle exoskeleton assistance on stability, which is currently incomplete.

Torque production at the ankle is an important contributor to balance, which suggests that ankle exoskeletons have the potential to improve stability during standing. Unimpaired adults tend to rely on modulating ankle torques to maintain a stable standing position, only switching to a hip strategy in response to large perturbations \cite{horak1986}. In older adults, maximum available plantar flexion and dorsiflexion torques are lower than in younger adults, reducing the range of perturbations that can be accommodated with an ankle strategy \cite{Hasson2014}. This reduction in maximum ankle torque (MT) has been shown in multiple studies to be correlated with decreased performance in balancing tasks \cite{Hasson2014} and a higher risk of falls \cite{Cattagni2014,CATTAGNI2016}. The rate at which this torque can be produced also declines with age \cite{Thelen1996}, although the effect of a lower maximum rate of torque development (MRTD) on standing balance is inconclusive. Reduced MRTD at the knee is associated with a history of falls \cite{bento2010, kamo2019}, while plantar flexor MRTD is significantly correlated with lower performance on a single standing leg balance test for older men, but not for older women \cite{ema2016}. Other studies have found that lower ankle MRTD is not a strong predictor of fall likelihood \cite{laroche2010, pijnappels2008}. Still, lower MRTD likely contributes, independently of lower MT, to increased fall risk in older adults \cite{pijnappels2008}.

Although ankle exoskeletons can, in principle, mitigate both MT and MRTD deficits by producing large amounts of torque quickly, the effect they have on stability is unclear. Depending on the population, device, controller, and perturbation paradigm, ankle exoskeletons have been reported to enhance stability \cite{Beck2023, Sharafi2023, RAO2006}, slightly reduce stability \cite{Canete2023, Sharafi2023, fang2023}, or have no effects on stability \cite{Son2010,Emmens2018}. Notably, the majority of balance studies for powered exoskeletons have focused on young and able-bodied users, who do not have impaired stability. As these devices are likely to be marketed to enhance mobility in older populations \cite{Raitor2024}, it is imperative to understand how they may affect the stability of users with age-related reduction in MT and MRTD. We currently do not know how these age-related changes may interact with additional exoskeleton assistance at the ankle. Because perturbation experiments on older adults are expensive, time-consuming, and risky, developing a generalizable model-based characterization of standing balance with ankle exoskeletons is a valuable, complementary approach. Modeling can provide insights into the mechanisms underlying exoskeleton-assisted standing balance, including how constraints on the human-exoskeleton system are affected both by the device and by age-related joint-level torque deficits.

Bounds on human stability for feet-in-place activities such as standing may be characterized by computing the set of all initial body center of mass (CoM) positions and velocities from which it is possible to stabilize to quiet standing, meaning that the final CoM position is above the foot and the CoM velocity is close to zero \cite{PAIPATTON1997}. This set is sometimes called the feasible stability region \cite{yang2009,bahari2021}.For a simplified linear inverted pendulum these are the well-known extrapolated center of mass boundaries (XCoM) \cite{HOF20051}. Herein we will use the term stabilizable region, abbreviated `SR' \cite{Raz2023}.

Stabilizable regions have been computed for various legged dynamical systems \cite{mummolo2017,orsolino2020}, including one example where the boundary was partially computed for a model of a human wearing an optimally controlled hip-knee-ankle exoskeleton \cite{Inkol2022}. The full stabilizable region under aging-related biological constraints and different ankle exoskeleton controllers has not been explored. Furthermore, formal stability properties of this region have not been defined. Invariance is of particular interest. If the SR is controlled invariant, this means that there always exists a controller that allows the human to stay within the region. If it is not, then there may be states within this region from which the human can pass through the SR, but not stabilize. Invariance guarantees that the SR does not include states which merely pass through the set. This property is especially important to verify when there are constraints on the control input. Such constraints may correspond to biologically meaningful quantities such as maximum joint torque, or to environmental constraints such as the frictional characteristics of the support surface.

We previously introduced a framework for determining the complete SR boundary for human dynamical systems \cite{Raz2023}. Our approach formulates the SR boundary as the solution of a single partial differential equation rather than a brute-force algorithmic search as in prior work, e.g. \cite{mummolo2017, Inkol2022}. Moreover we provide strong formal invariance and stability guarantees over the entire SR in the presence of both biological and environmental constraints.

Here we present a generalizable, model-based understanding of how an ankle exoskeleton may alter the domain of invariant stable motions available to a user in the presence of joint level functional changes associated with aging, including reduced MT production and MRTD. To our knowledge, this is the first study that attempts to systematically understand the combined effect of ankle exoskeleton assistance and age-related ankle torque deficits on feasible stability. Using common exoskeleton control strategies, we analyze the effect that exoskeletons have on the invariant stabilizable region for models of young and old adults. We show that while exoskeletons can indeed enhance ankle torque production and improve stability under certain conditions, they can also act as a disturbance, hindering stability in safety-critical regions of the state space.

\section{Preliminaries} \label{sec:preliminaries}
Our methods for computing stabilizable regions (SRs) are based on two notions, controlled invariance and backward reachability, which we formally define in this section.
\subsection{Controlled invariance and backward reachability}
\begin{figure}[ht] 
    \centering
    \includegraphics[width=0.45\textwidth]{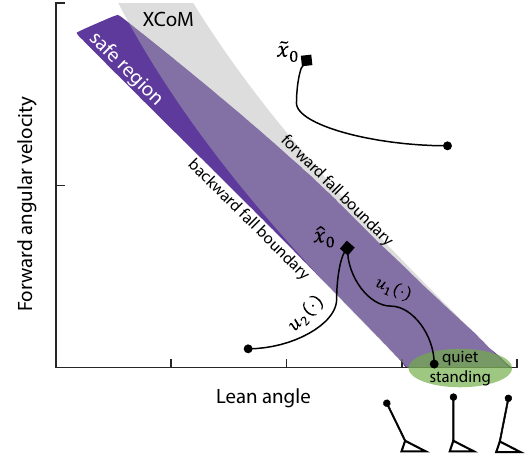}
    \caption{Illustration of a target set (green) and the set of all states from which it can be reached. This set is called the backward reachable set (BRS, in purple). The state $\hat{x}_0$ is in the BRS, because there exists at least one controller, $u_1$, that drives the system to the target set. No such controller exists for $\Tilde{x}_0$ so it is not in the BRS. If the target set is controlled invariant then the union of the target set and the BRS is also invariant. Here, the BRS was computed with model parameters corresponding to a young male, using our method described in Section \ref{sec:methods}. The gray region is computed using the extrapolated center of mass (XCoM) method from \protect\cite{HOF20051}, which we have converted here to angular coordinates.}
    \label{fig:BRS}
\end{figure}
Consider the system
\begin{equation} \label{generecNonLSys}
    \Dot{x} = f(x,u),
\end{equation}
where state $x \in X \subset \mathbb{R}^n$, input $u \in U \subset \mathbb{R}^m$. Control signals are denoted $u(\cdot) \in \mathcal{U} = \{\phi : [0, \infty] \rightarrow U\}$. Solutions to this system are functions of time, and we denote them by $\varphi(\cdot;x_0, u(\cdot))$, for initial condition $x_0$ and control signal $u(\cdot)$.

A set $\Omega \subset X$ is controlled invariant if, for any initial condition in the set, there exists a controller that maintains the system state within $\Omega$:
\begin{definition}
A set $\Omega \subset X$ is \textit{ controlled invariant} for system \eqref{generecNonLSys} if, for all $ x_0 \in \Omega,$ there exists $u(\cdot) \in \mathcal{U}$ such that for all $ t \in [0, \infty]$, $\varphi(t;x_0, u(\cdot)) \in \Omega$. 
\end{definition}
 If $\Omega$ is a safe region in the state space, then the ability to always stay within the set is a desirable property.

We can also define the region of the state space from which it is possible to reach a specified (not necessarily invariant) subset, $S$. The set of all $x \in X$ such that there exists a control signal that drives the system to $S$ within some finite time horizon is the backward reachable set of $S$. 

\begin{definition} Let $S \subset X$ and $T \in \mathbb{R}_0^+$. Then $\mathcal{G}_T(S)$, the \textit{backward reachable set} of $S$ at time $T$, is
\begin{equation} \label{def:BRS}
    \mathcal{G}_T(S) := \{ x \in X \mid \exists u(\cdot) \in \mathcal{U} \, s.t. \, \varphi(T;x,u)  \in S \}. 
\end{equation}
\end{definition}
An illustration of this set is depicted in Figure \ref{fig:BRS}. Because this definition does not constrain the form of the control law, it is the \textit{maximal} backward reachable set (BRS) with respect to $\mathcal{U}$. Thus it represents the `best case' scenario of the states from which it is possible to reach the target set. 

The BRS contains states from which the system can reach the target set at precisely time $T$. The set of states from which the target set can be reached at $t \leq T$ is called the backward reachable tube, and is a union of backward reachable sets:
\begin{definition}
The \emph{maximal backward reachable tube} of a set $S\subset X$ over a time horizon $[0,T]$ is $\mathcal{G}_{[0,T]}(S) = \cup_{t\in[0,T]}\mathcal{G}_t(S)$.
\end{definition}

Given these definitions, we note the following two facts:
\begin{fact}
If $\Omega$ is controlled invariant, then its backward reachable set is  controlled invariant. 
\end{fact}
\begin{fact}

Assume a system as defined in \eqref{generecNonLSys}. Let $S \subset X$ be controlled invariant. Then $\mathcal{G}_{t_1}(S) \subset \mathcal{G}_{t_2}(S)$ for all $t_1$, $t_2\in \mathbb{R}^+$ with $t_1 < t_2$. 
\end{fact}
Fact 2 is proved in \cite{Raz2023}. Taken together, these facts mean that for an invariant target set $S$,  $G_{[0,T]}(S) = G_T(S)$. Thus it is sufficient to compute the  backward reachable set rather than computing the full reachable tube.  

\subsection{Relating invariance and reachability to human stability}
The relationship between human stability and our mathematical definitions is illustrated in Figure \ref{fig:BRS}.
Assume that the dynamics \eqref{generecNonLSys} represent a parameterized model of a standing human.
In the figure, the green ellipse represents a controlled invariant target set of states corresponding to quiet standing. For the purposes of the illustration the set is depicted to be larger than it is.
The purple region is the true-to-size backward reachable set that we compute using a constrained dynamical model with parameters corresponding to a young male, as described later in Section \ref{sec:model}.
The boundaries of this set correspond to forward and backward fall boundaries, and to constraint failures. 
Outside of these boundaries the ankle strategy fails and the model must take some action to avoid a fall, such as taking a step. For visual clarity we only depict the portion of the state space corresponding to forward angular velocities.  
Because the target set is controlled invariant, the resulting purple safe region is as well.
For comparison, we show the same bounds computed using the XCoM from \cite{HOF20051}, in light gray. XCoM is often used by biomechanists to estimate dynamic stability boundaries for humans as they stand and walk \cite{watson2021}.

To determine the stability bounds of the human-exoskeleton system, we compute reachable sets with the dynamics of an ankle exoskeleton controller added to \eqref{generecNonLSys}, the details of which are presented in Section \ref{sec:model} and \ref{sec:methods}. 
We construct invariant target sets corresponding to quiet standing with exoskeleton assistance, and then compute the backward reachable sets. 
The resulting sets then delineate the feasible stabilizable boundaries with ankle exoskeleton assistance. 

\section{Human-Exoskeleton Model}
\label{sec:model}

\begin{figure}[h] 
    \centering
    \includegraphics[width=0.45\textwidth]{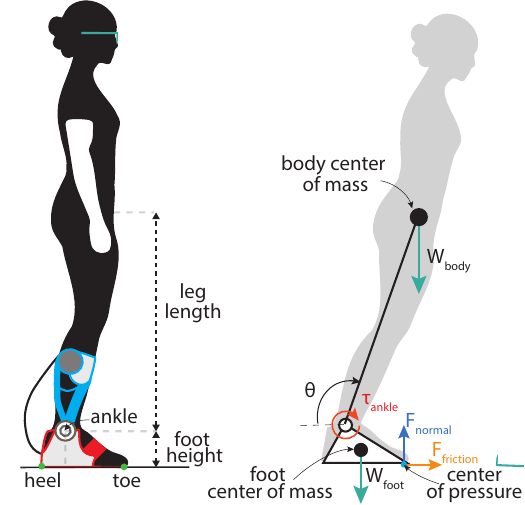}
    \caption{Free body diagram of the combined human-exoskeleton system. 
}
    \label{fig:FBD}
\end{figure}
We use a planar, two link, single degree-of-freedom model to represent a standing human (Fig  \ref{fig:FBD}). The first link represents the foot, while the second link represents the rest of the body. These links are connected using a pin-joint at the ankle, with the axis of rotation lying normal to the sagittal plane. The mass of each link is assumed to be lumped at the center-of-mass for the respective body. The human interacts with the ground through forces applied on the foot segment. Thus, these ground reaction forces applied at the foot, and the weight of the foot and body segments applied at each respective mass center, are the only external forces acting on the human system.

We model the ankle exoskeleton as a massless device, worn bilaterally, that only generates torque in the sagittal plane. We assume that the human can resist the torque generated by the device if necessary. Thus the exoskeleton is represented by an ideal torque actuator at the ankle, and the net torque at the ankle is the sum of the human and the exoskeleton torques. 

Rather than including the human torque $\tau_{\text{human}}$ as an ideal (instantaneous) torque actuator, we use an actuation model where the direct human input to the system is the desired rate of torque development, $\Dot{\tau}_{\text{human}}^{\text{des}}$. The stabilizable configurations are thus represented in joint angle-velocity-torque space, where the torque state represents the human torque only. Denoting the state space as $X \subset \mathbb{R}^3$, with $x_1 = \theta, x_2 = \Dot{\theta}$ and $x_3 = \tau_{\text{human}}$, and the input space as $U \subset \mathbb{R}$ with input $u = \Dot{\tau}_{\text{human}}^{\text{des}}$, in state space form, our dynamics are

\begin{equation} \label{eq:Lifted_IP_dynamics}
    \begin{bmatrix}
    \Dot{x}_1\\
    \Dot{x}_2\\
    \Dot{x}_3
    \end{bmatrix} = 
    \begin{bmatrix}
    x_2\\
    -\frac{g}{l}\cos{x_1} - \frac{b}{ml^2}x_2 + \frac{1}{ml^2}(x_3 + \tau_{\text{exo}})\\
    u
    \end{bmatrix}.
\end{equation}
The constants $g$, $m$, and $l$ represent the gravitational acceleration, mass of the body link, and leg length. The parameter $b$ represents minimal damping that is included to make the dynamics less stiff.

We can thus limit the human MT and MRTD of the model by constraining state $x_3$ and input $u$:
\begin{align}
& \tau_{\text{human}}^{\text{max pf}} \leq x_3 \leq \tau_{\text{human}}^{\text{max df}}& \tag{C1}
\label{const:torqueBds}\\
& \text{MRTD}^{\text{max pf}} \leq u \leq \text{MRTD}^{\text{max df}}& \tag{C2}
\label{const:RTDBds}
\end{align}
Maximum plantar flexion torque $\tau_{\text{human}}^{\text{max pf}}$ is negative in our angle convention, while $\tau_{\text{human}}^{\text{max df}}$ denotes positive dorsiflexion torque. Lower magnitudes of torque bounds correspond to weaker ankle strength, while lower bounds on $u$ correspond to reduced rate of torque development, as observed in older 
adults \cite{Thelen1996}.
 
Foot contact with the ground is determined by the normal force $F_\text{normal}$, frictional force $F_\text{friction}$ with coefficient of static friction $\mu$, and the location of the center of pressure (CoP). The CoP should remain within the base of support (BoS), which corresponds to the length of the foot. Full ground contact is enforced by three zero-moment point (ZMP) safety constraints which prevent the foot from tipping or slipping \cite{vukobratovic2004}:
\begin{align}
& F_\text{normal} \geq 0& \tag{C3}
\label{GRFconstraint}\\
& \text{ CoP} \in \text{BoS},&
\label{CoPconstraint}\tag{C4}\\
& \vert F_\text{friction} \vert < \mu F_\text{normal}.&
\label{frictionConstraint}\tag{C5}
\end{align}

A human user could respond to the assistive torque of the exoskeleton in a range of ways. They may actively resist the torque generated by the device, or they may adapt their torque production to allow the exoskeleton to assist as much as possible. Studies with able-bodied participants, such as \cite{Emmens2018} have shown that young adults adapt quickly to an exoskeleton using a state-feedback controller, with minimal alteration to CoM kinematics. Thus we make the following assumptions related to the human-exoskeleton interaction:
\begin{enumerate}
    \item The exoskeleton employs a state feedback control strategy and is unaware of the foot-ground interaction constraints. It cannot reduce or increase torque production if a constraint is in danger of being violated
    \item The human adapts their torque production to account for the torque produced by the exoskeleton
    \item The human cannot instantaneously change the value of the torque produced at the ankle
\end{enumerate}

Assumption 3 is addressed via constraint \eqref{const:RTDBds}.
Because of Assumption 2, \eqref{GRFconstraint}-\eqref{frictionConstraint} can be expressed as constraints on the total torque of the system such that the exoskeleton torque is accounted for. Let $h: X \rightarrow \mathbb{R}.$ Then we can write constraints on the torque as:
\begin{equation} \label{eq:torqueConstraintswithExo}
    h_\text{lb}^k(x) < x_3 + \tau_\text{exo}(x) < h_\text{ub}^k(x)
\end{equation}
for $k \in \{1, 2, 3\}$. As we assume the exoskeleton torque is a function of state, these are actually state constraints on the human torque state $x_3$. Appendix I of the supplementary material includes a full derivation of the equations of motion and foot-ground contact constraints.

\section{Methods}
\label{sec:methods}
We assess the effect of ankle exoskeletons on feasible stability by computing the BRS of the human-exoskeleton system for two exoskeleton controllers, over a range of sex- and age-adjusted model parameters.
\subsection{HJB Reachability}
    The BRS can be computed using Hamilton Jacobi reachability tools. Hamilton Jacobi Bellman reachability uses HJB partial differential equations to compute implicit set representations of the backward reachable set \cite{Mitchell2005}. We provide a brief overview of the method here, but we note that our analysis can be done with any nonlinear continuous time backward reachability tool. A more rigorous development of HJB and its use for solving reachability problems can be found in \cite{Mo2018}.
Given a dynamical system of the form \eqref{generecNonLSys} and a target set $S \subset X$, let $g(x)$ be a function whose zero sublevel set represents $S$, i.e.,  $S= \{ x\mid g(x) \leq 0\}$. Typically $g(x)$ is a signed distance function. HJB reachability tools compute the backward reachable set by solving the PDE
\begin{equation} \label{HJBPDE}
    D_t V(t,x) + H(t,x, \nabla V) = 0,\, V(0,x) = g(x).
\end{equation}
over reversed time interval $[-T, 0]$, where $T$ is the desired duration and $t \in [-T, 0]$.  The solution $V(t,x)$ is called the value function. 
The function $H(t,x, \nabla V)$ is called the Hamiltonian, and in this case is given by
\begin{equation} \label{hamfunc}
    H(t,x, \nabla V)  = \min_{u\in U} \nabla V \cdot f(x,u)
\end{equation}
The argument of the Hamiltonian,
\begin{equation} \label{optcontrol}
    a^*(x) \in \arg \min_{u\in U} \nabla V \cdot f(x,u)
\end{equation}
is an optimal control input at a given state $x$ that minimizes $V(x,t)$ along the system trajectory. 
The value function itself is an implicit surface function representation of the backward reachable set at time $t$, $\mathcal{G}_t$, as defined in Definition \ref{def:BRS}. Thus,  
\begin{equation}
    \mathcal{G}_t = \{ x\mid V(t,x) \leq 0\}.
\end{equation}
Note that we abuse notation slightly by letting $t$ also denote $t \in [0,T]$ in the nonreversed time interval.

Suppose now that system \eqref{generecNonLSys} is control affine, such that it can be written
\begin{equation}\label{controlaffinesys}
    \dot{x} = f(x,u) = f^x(x) + f^u(x)u.
\end{equation}
We assume that $u \in U \subset \mathbb{R}^m$ and that $U$ is a hyperrectangle such that $u$ is bounded elementwise, i.e. $u_i \in [\underline{u_i}, \overline{u_i}]$. In this case, a simple analytical solution to Problem \eqref{optcontrol} exists. As the optimization is over $u$, portions of the resulting expression that do not contain the control input can be ignored, and the problem reduces to

\begin{equation} \label{eq:optCtrlSum}
   a^* \in \arg\min_{u\in U}\sum_{i=1}^{n} \frac{\partial V(t,x)}{\partial x_i} f^u_i(x) \cdot u_i
\end{equation}
The optimization problem is evaluated at a fixed time $t$ and state $x$, which means that each coefficient $\frac{\partial V(t,x)}{\partial x_i} \cdot f^u_i(x)$ is a constant. The solution can thus be found by letting $u_i = \underline{u_i}$ if the coefficient is positive and $u_i = \overline{u_i}$ otherwise.
For system \eqref{eq:Lifted_IP_dynamics}, equation \eqref{eq:optCtrlSum} reduces to
\begin{equation}
    \min_{u \in U} H(t, x, \nabla V) = \frac{\partial V(x,t)}{\partial x_3}u.
\end{equation}
Assuming $U = [\underline{u}, \overline{u}]$, the optimal controller is therefore $\overline{u}$ if coefficient $\frac{\partial V(x,t)}{\partial x_3} < 0$ and $\underline{u}$ otherwise. 
\subsection{Constructing invariant target sets} \label{subsec:ConstructTS}
Recall that controlled invariance is a desirable safety property. To ensure that the BRS is controlled invariant, it suffices to show that the target set is controlled invariant (Section \ref{sec:preliminaries}). However, in our case, the target set must be a subset of the safe region defined by constraints \eqref{const:torqueBds} and \eqref{GRFconstraint}-\eqref{frictionConstraint}, and obey constraints on the controller \eqref{const:RTDBds}. Thus the target sets must be constructed with care.

Appropriate candidates for the target set can be found in the biomechanics literature, where feasible stability is commonly defined as the ability to reach static standing configurations \cite{PAIPATTON1997}. These are equilibrium configurations where (i) the angular velocity of the body is zero and (ii) the $x$-coordinate of the center of mass ($\text{CoM}_x$) is within the base of support (i.e. above the foot). These configurations form a continuum, a subset of which can be used to construct target sets.

We first analyze static equilibrium (zero net-torque) states that fulfill condition (i). 
Let $\tau_{\text{act}}$ denote the total actuation torque about the ankle, as in Figure \ref{fig:FBD}, and let $\tau_g = mgl\cos{\theta}$ be the gravitational torque. Summing the torques, we see that $\tau_{\text{act}} = \tau_g$ meaning that the net actuation torque must remain constant and compensate for gravity.
With human torque $\tau_{\text{human}}$ state feedback exoskeleton torque $\tau_{\text{exo}}$, we have that $\tau_{\text{act}} = \tau_{\text{human}} + \tau_{\text{exo}}$. Also recall that in our `lifted' system \eqref{eq:Lifted_IP_dynamics}, input $u = \Dot{\tau}_{\text{human}}$, is the rate of the human torque development. 
Therefore, at equilibrium, $\tau_{\text{human}} = \tau_g - \tau_{\text{exo}}$ and $u = 0$. 

\begin{figure}[h] 
    \centering
    \includegraphics[width=0.45\textwidth]{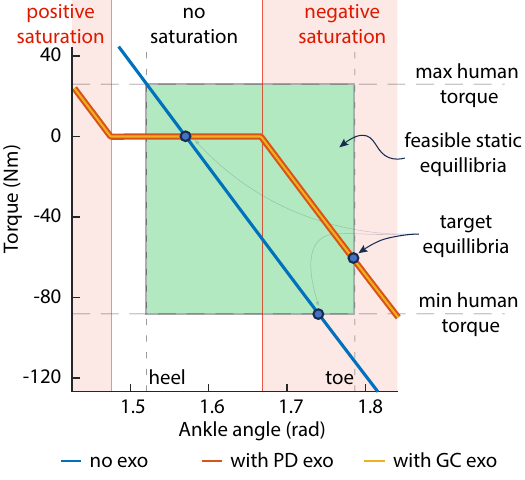} 
    \caption{Continua of equilibrium points in the zero-velocity plane for the older female model with and without an exoskeleton. The shaded green region indicates the bounds of the foot along the horizontal axis, and the human torque bounds on the vertical axis. To generate sets representing quiet standing, we construct ellipsoids centered at the ankle and at the point closest to the toe corresponding to maximum feasible static support. These points are marked with blue circles. 
    }
    \label{fig:targetSets}
\end{figure}

There are multiple static equilibria corresponding to various resting states where the net actuation torque counters gravity, which can be defined as the following set
\begin{equation}\label{eq:staticEqSet}
    \Omega_{eq} = \{x \in X \vert x_2 = 0, x_3 = \tau_g - \tau_{\text{exo}}\}.
\end{equation} 
Curves representing these equilibria for system \eqref{eq:Lifted_IP_dynamics} with and without an exoskeleton are shown in Figure \ref{fig:targetSets}. The depicted exoskeleton controllers are described in more detail in the following sections and saturate at $\SI{50}{N\cdot m}$.

Our desired static equilibria must also satisfy condition (ii), with the $\text{CoM}_x$ above the foot. In angular coordinates, the bounds of the foot can be represented as the maximum forward and backward lean angle, such that the $\text{CoM}_x$ is within the foot. We refer to the maximum forward lean as $\theta_{\text{toe}}$ and the backward lean angle as $\theta_{\text{heel}}$. Using the geometric parameters of the foot as shown in Figure \ref{fig:FBD}, we have that $\theta_{\text{heel}} = \cos^{-1}{(\frac{a}{l}})$ and $\theta_{\text{toe}} = \cos^{-1}{(\frac{l_f -a}{l}})$. This gives a range of permissible target lean angles, $\Theta_{\text{range}} = [\theta_{\text{heel}}, \theta_{\text{toe}}]$. 

Lastly, we account for the limits on $\tau_\text{human}$. Feasible equilibria must lie within the intersecton of $\Theta_{\text{range}}$ and constraint \eqref{const:torqueBds} (the green square in Figure \ref{fig:targetSets}). 
Note that static equilibria within this square also satisfy constraints \eqref{GRFconstraint}-\eqref{frictionConstraint}. At these states, the system is not accelerating and is at zero velocity - therefore the ground reaction force is equal to the weight of the foot and body, there is no friction force, and the CoP is within the bounds of the foot. 
Thus a possible target set is the intersection of $\Omega_{eq}$ with the safe region, where the foot maintains contact with the ground without tipping or slipping (Fig. \ref{fig:targetSets}, orange/blue curves inside green square). 

This representation of the target set raises two key concerns. First, because the set consists only of static equilibria, it does not allow for natural postural sway, which is a fundamental characteristic quiet standing in humans \cite{ERA2006}. Secondly, one-dimensional target sets consisting of curves or line segments can lead to computational issues when computing the BRS.
To generate target sets that are more biologically realistic, we instead use a Control Lyapunov Function (CLF) to construct controlled invariant ellipsoids. The resulting ellipsoidal target sets encompass a range of realistic sway velocities and are large enough to avoid problems with computational precision. 

CLFs are an extension of Lyapunov functions to systems with control, and are defined as follows \cite{sontag1983}:
\begin{definition} \label{def:CLF}
  Let $x^* = 0$ be an equilibrium point of \eqref{generecNonLSys} and let $E: \mathbb{R}^n \rightarrow \mathbb{R}$ be a positive-definite, continuously differentiable and radially unbounded function. Then  $E(\cdot)$ is a \textit{control Lyapunov function} if for all $x \neq 0$ there exists $u$ such that $\Dot{E}(x,u) < 0$.
\end{definition} 
In other words, a control input exists that stabilizes the system at $x^*$. Importantly, just as the level sets of Lyapunov functions are invariant, the level sets of a CLF are controlled invariant. CLFs can therefore be used to construct a target set by (i) selecting equilibrium points, (ii) computing CLFs about those points, (iii) determining the appropriate set level and (iv) taking the union of the resulting level sets as the target set.

We start by selecting relevant equilibrium points from \eqref{eq:staticEqSet}.
When the system includes a saturating exoskeleton torque due to device motor limits, the dynamics become hybrid. There are three modes corresponding to 1) negative saturation, 2) no saturation, and 3) positive saturation. 
Only two modes appear within the green region, negative saturation and no saturation (Fig. \ref{fig:targetSets}). Thus, we choose one point per mode: a `toe' point at negative saturation, and an `ankle' point at no saturation.

To select the equilibrium point at the toe, we must take into account the functional base of support (FBOS). The FBOS is defined as the maximum range that an individual can voluntarily translate their CoP in the anterior (toward the toe) and posterior (toward the heel) directions. In our model, we assume that the angle corresponding to the maximum anterior CoP position is approximately the most anterior point contained within the foot and maximum plantar flexion torque bounds introduced in constraint \ref{const:torqueBds}:
\begin{equation}
    \theta_{\text{toeEq}} = \max{(\theta_{\text{toe}},\cos^{-1}(\frac{\tau_{\text{human}}^{\text{max pf}} - \tau_{\text{exo}}^{\text{sat}}}{mgl}))} - 0.03.
\end{equation}

The angle $\theta_{\text{toeEq}}$ is the largest angle such that the model's human and exoskeleton maximum torques,$\tau_{\text{human}}^{\text{max pf}}$ and $\tau_{\text{exo}}^{\text{sat}}$, are sufficient to compensate for gravity and maintain a static position. We subtract $0.03$ \unit{rad} to allow for displacement due to postural sway. When the model torque bounds correspond to a young individual this point occurs at $\theta_{\text{toe}}$, but in models of older adults, the region of static support may be smaller. Indeed, in an older adult there may be further limiting factors shrinking the FBOS, such as reduced toe flexor strength \cite{Endo2002}. 

The equilibrium point corresponding to the toe is therefore $x_{\text{toeEq}} = [\theta_{\text{toeEq}}, 0, mgl\cos{(\theta_{\text{toeEq}}) - \tau_{\text{exo}}^{\text{sat}}}]$ (Fig. \ref{fig:targetSets}). 
 The ankle equilibrium point is more straightforward, corresponding to upright standing with no lean:
  $x_{\text{ankEq}} = [\theta_{\text{ank}}, 0 , \tau_g - \tau_{\text{exo}}]$.
 Here $\tau_{\text{exo}}$ is determined by a state feedback law $K_{\text{exo}}(x_1, x_2)$.  
 Note that in the no exoskeleton case, $\tau_{\text{exo}} = \tau_{\text{exo}}^{\text{sat}} = 0$.

\begin{figure*}[h] 
    \centering
    \includegraphics[width=\textwidth]{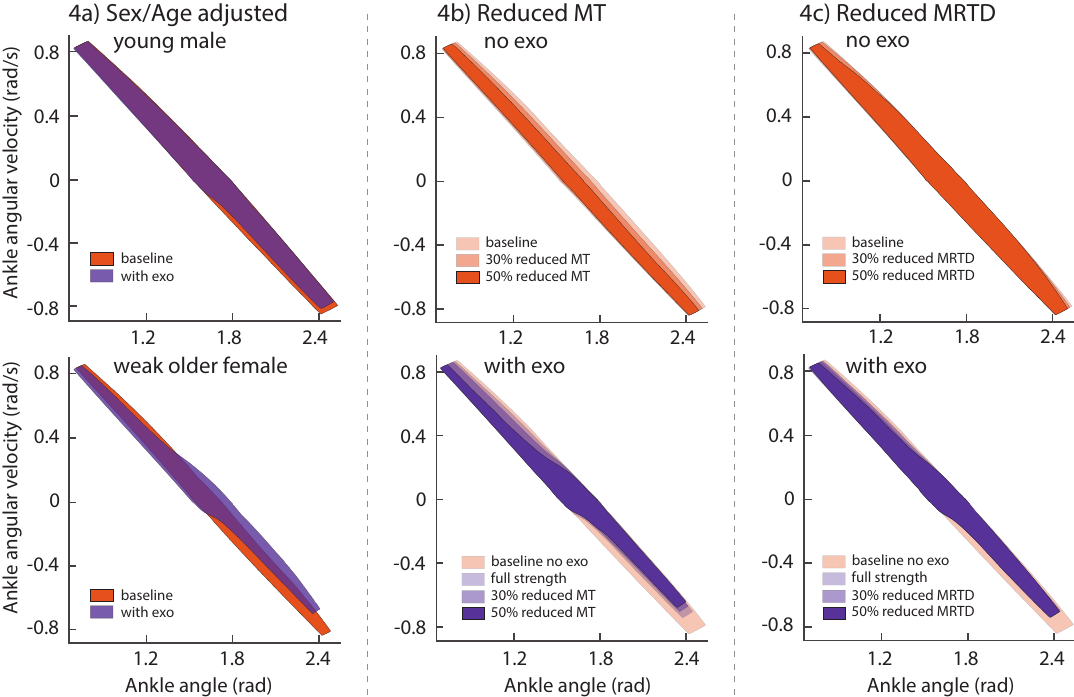}
    \caption{Panel (4a) shows the stabilizable region computed for the young male model (top) and weak older female (bottom). The baseline regions without the exoskeleton are shown in orange, while the region with exoskeleton assistance is shown in purple. Stabilizable regions when MT and MRTD are reduced independently are shown in the panels (4b) and (4c). In the top row, stabilizable regions are computed without an exoskeleton. The largest sets are baseline, followed by 30\% reduction and 50\% reduction in each respective quantity. In the bottom row, the stabilizable regions with exoskeleton assistance are shown in purple, while the reference baseline (full strength, without exoskeleton) is shown in light orange.}
    \label{fig:SixPanel}
\end{figure*}

We next construct CLFs by linearizing \eqref{eq:Lifted_IP_dynamics} for each equilibrium point $x_{\text{eq}} \in \{x_{\text{ankleEq}}, x_{\text{toeEq}}\}$:
\begin{equation} \label{eq:linSys}
    \Dot{\Tilde{x}} = A\Tilde{x} + B\Tilde{u}
\end{equation}
where $\Tilde{x} = x - x_\text{eq}$ is the new state variable representing deviation from $x_\text{eq}$.
We can find a CLF of the form $E(\Tilde{x}) = \Tilde{x}^TP\Tilde{x}$ with a corresponding stabilizing human controller $\Tilde{u} = K(\Tilde{x})$ by solving a semidefinite program, described in Appendix II-A in the supplementary material.
Controlled invariant ellipsoids centered at each $x_\text{eq}$ can then be formulated as $\{x \mid (x-x_\text{eq})^TP(x-x_\text{eq}) \leq c\} $ for some $c \in \mathbb{R}$, with the union of these sets becoming the target set. Note that as each CLF is valid for the unconstrained linearized system \eqref{eq:linSys}, each $c$ should be chosen small enough such that the state and input constraints on the original nonlinear system are satisfied for all $x$ in the ellipsoid. Our method for computing appropriately scaled ellipsoids is described in detail in Appendix II-B of the supplementary material.

\subsection{Evaluating the effect of ankle exoskeleton assistance}
To determine how exoskeleton assistance may affect the stability of older adults with differing ankle torque production capabilities, we model five representative individuals. The individuals correspond to the young female (YF), young male (YM), older female (OF), and older male (OM) categories described in \cite{Thelen1996}. 
We create the fifth category, weak older female (wOF), by reducing the maximum isometric torque and normalized rate of torque development by half a standard deviation of the mean values for the OF in \cite{Thelen1996}. 
The respective MT and MRTD bounds of each model are shown in Table \ref{table:model_params}.

For each model, we construct appropriate quiet standing target sets (section \ref{subsec:ConstructTS}) and use a state of the art HJB toolbox\cite{Mitchell2005} to compute the backward reachable set in position-velocity-torque space.
To form the stabilizable region, we project the resulting BRS into position-velocity space. 
We first compute a baseline SR for a young male model without exoskeleton assistance (Fig. \ref{fig:BRS}, purple).
We validate that for a full strength YM model, the bounds approximately coincide with those computed via the XCoM method (Fig. \ref{fig:BRS}, gray). These bounds overlap wherever the model is close to linear. 

We then compute the BRS with bilateral ankle exoskeleton assistance added. We assume that the motor on each exoskeleton saturates at $\SI{25}{N\cdot m}$ for a total of $\SI{50}{N\cdot m}$ and that each device uses a gravity compensation (GC) control strategy, as described in \cite{Emmens2018}:
\begin{equation} \label{eq:gravComp}
K(x) = 
\begin{cases}
    m g l \cos{\theta} & \text{if } \abs{\tau_{\text{exo}}} < \text{motor limit}\\
    \pm \text{motor limit} & \text{otherwise},
\end{cases}
\end{equation}

To understand the effect of damping, we also compute the SR with a proportional derivative (PD) controller where the desired setpoint is static, upright standing:
\begin{equation}
K(x) = 
\begin{cases}
    K_p(\theta - \frac{\pi}{2}) + K_d(\Dot{\theta}) & \text{if } \abs{\tau_{\text{exo}}} < \text{motor limit}\\
    \pm \text{motor limit} & \text{otherwise}
\end{cases}
\end{equation}
where $K_p, K_d > 0$ represent the PD controller gains. The gains are also taken from \cite{Emmens2018}, where proportional gain $K_p = mgl$ is a linearized gravity compensation term, while the derivative gain $K_d = 0.3\sqrt{ml^2K_p}$ compensates for gravitational stiffness. Both the GC and PD controller are easily substituted into the dynamics such that $\tau_{\text{exo}} = K(x)$.

For a high-level understanding of the effect of each control strategy, we compute the total area of the SR without exoskeleton to form a baseline for backward and forward ankle angular velocities. We then determine the total area of the human-exoskeleton SR computed with both GC and PD controller strategies, and calculate the percent change in total area with respect to the baseline. While we analyze two strategies, we emphasize that the methods presented here can be applied to any state-feedback controller.

The three older models in Table \ref{table:model_params} combine reductions in MT and MRTD. To get a clearer picture of the independent contributions of reduced MT and MRTD, we also analyze how reduced MT and reduced MRTD affect the SR separately. We compute the SR for the YF model with the MRTD held constant at the YF's nominal peak value, while reducing the MT in ten percent increments. We repeat this procedure reducing MRTD while holding MT fixed. We evaluate how deficits in MT and MRTD interact with ankle exoskeleton assistance by computing similarly reduced strength sets with GC exoskeleton assistance added. We compute the percent reduction or increase in area of the  stabilizable regions for the reduced-strength vs. full-strength models.

\begin{table}[h!]
\centering
\caption{Model parameters for young female (YF), old female (OF), young male (YM), old male (OM), and weak older female (wOF). Including maximum torque (MT) and maximum rate of torque development (MRTD) in dorsiflexion (DF) and plantar flexion (PF) directions}
\begin{tabular}{|l||l c c c c c||} 
 \hline
 & Dir. & YF & OF & YM & OM & wOF  \\ [0.5ex] 
 \hline\hline 
 Mass (kg) & n/a & $59.4$ & $60.0$ & $72.9$ & $74.5$  & $60.0$ \\
 Height (m) & n/a & $1.65$ & $1.59$ & $1.77$ & $1.74$ & $1.59$ \\
 MT (Nm) & DF & $\phantom{0}28$ & $\phantom{0}22$ & $\phantom{0}43$ & $\phantom{0}37$ &
 $\phantom{0}21$\\
    &  PF  & $130$ & $\phantom{0}88$ & $181$ & $137$ & $\phantom{0}78$ \\
 MRTD (Nm/sec) & DF & $219$ & $148$ & $309$ & $232$ & $130$\\  
 & PF & $608$  & $389$ & $957$ & $681$ & $303$\\ [1ex] 
 \hline
\end{tabular}
\label{table:model_params}
\end{table}

\subsection{Work-energy analysis} \label{Methods:Work-energy}
We perform a work-energy analysis of a simplified model with and without the GC controller to complement our backward reachability analysis. While less accurate, this allows us to better understand the physical and physiological mechanisms that lead to changes in the SR between the no-exoskeleton and exoskeleton conditions.
We assume conservation of energy and use unconstrained and undamped 2-D pendulum dynamics, where $\tau_h + \tau_\text{exo} = ml^2\Ddot{\theta} + mgl\cos(\theta)$. 

We observe that gravity exerts a negative torque when the CoM is behind the ankle and a positive torque when the CoM is ahead of the ankle. The `zero gravity torque' line at $y = \theta_\text{ankle}$ separates the 2d state space into positive and negative gravity torque regions (dashed vertical black line in Fig. \ref{fig:discussionFig}a). We also define a \textit{zero human-torque line}:

\begin{definition}
    The \textit{zero human-torque line (ZTL)} is a line in the phase plane along which the human does not need to produce torque to reach a final static standing condition.  
\end{definition}
Consider first, the no exoskeleton case --- The only static standing position where the human does not need to produce a gravity resisting torque is directly above the ankle. Thus, there is a unique ZTL along which total mechanical energy is constant, being equal to the potential energy when the center of mass is directly above the ankle. If the initial energy is $E_i = mgl\sin{\theta} + \frac{1}{2}ml^2\dot{\theta}^2$ and the final energy is $E_f = mgl\sin{\theta_{\text{ankle}}}$ we can compute the ZTL by letting $E_i = E_f$, and derive an equation for the system velocity as a function of position: 
\begin{equation} \label{eq:ZHTT_noExo}
    \dot{\theta} = \pm \sqrt{2\frac{g}{l}(\sin{\theta_{\text{ankle}}} - \sin{\theta})}
\end{equation} 

This ZTL (Fig. \ref{fig:discussionFig}a) delineates the direction of the optimal human torque.
For a trajectory starting below the ZTL, a positive (dorsiflexion) torque must eventually be applied to avoid a backward fall past the heel. 
For trajectories starting above the zero-torque line, a negative (plantar flexion) torque is eventually necessary to brake quickly enough to prevent the CoM from overshooting past the toe.
Thus, the boundary of the SR that is below the ZTL corresponds to the lower bound before a backward fall occurs, whereas the SR boundary above the ZTL corresponds to forward falls \cite{PAIPATTON1997}.
Based on these fall boundaries, the torque applied by gravity works to prevent falls rather than cause them, in large portions of the state space.

Now we consider how the addition of an exoskeleton with GC control changes the ZTL. Equation \eqref{eq:ZHTT_noExo} is no longer valid once a 
GC torque is added, as the exoskeleton will  apply a torque everywhere except when the gravitational torque is zero. To follow the same trajectory as the original ZTL, i.e. without an exoskeleton, the human would have to exactly counter the exoskeleton torque. Thus, the ZTL changes in the presence of the exoskeleton, and needs to be re-calculated.

Recalling that the exoskeleton torque saturates at specific negative and positive values, we can calculate the angles at which saturations occur. We divide the state space into three regions: with positive saturation, where $\tau_\text{exo} = \tau_\text{sat}$; no saturation, where $\tau_\text{exo} = mgl\cos{\theta}$; and negative saturation,where $\tau_\text{exo} = -\tau_\text{sat}$. The ZTL is computed piecewise for each region.

In the region where the controller is not saturated, the exoskeleton torque fully compensates for the gravitational torque, meaning that the human must generate no additional torque along this range of angles. The endpoints of this range are where the exoskeleton saturation torque can fully compensate for gravity without additional torque from the human. Denoting these angles as $\theta_\text{sat}^\text{ub}$ and $\theta_\text{sat}^\text{lb}$, the ZTL can be derived using the work-energy principle, similar to \eqref{eq:ZHTT_noExo} but with the addition of the work due to the saturated exoskeleton. For example, 
\begin{equation} \label{eq:ZHTT_withExo}
    \dot{\theta} = \sqrt{2(\tau_{\text{exo}}^{\text{sat}}(\theta_\text{sat}^\text{ub} - \theta) + \frac{g}{l}(\sin{\theta_\text{sat}^\text{ub} - \sin{\theta}))}}
\end{equation}
is the ZTL in the region of positive exoskeleton saturation. The full ZTL with exoskeleton assistance is shown in Figure 6a as a solid purple line.

We can use the work-energy principle to characterize how exoskeleton-induced changes to the ZTL, combined with the human torque bounds, affect stabilizability. The net work of the system is the sum of the work done by the human, exoskeleton, and gravity.
Note that the net work done by gravity and the exoskeleton will always have opposite signs due to the controller formulation. For most initial conditions reaching the target set requires a reduction in kinetic energy, i.e. negative work,
\begin{equation}
    -ml^2\dot{\theta}_i^2 = \int_{\theta_i}^{\theta_f} (\tau_\text{exo} + \tau_\text{gravity} + \tau_\text{human}) d\theta .
\end{equation}

 By selecting an appropriate final position $\theta_f$, and by assuming that the human produces a constant optimal torque, upper and lower boundaries on the stabilizable region can be computed in a piecewise manner in the position-velocity plane. For example, for an initial angle $\theta_i < \theta_\text{sat}^\text{lb}$, the upper boundary is expressed by the relation
\begin{equation}
\begin{aligned}
    -ml^2\dot{\theta}_i^2 =& \tau_{\text{exo}}^{\text{sat}}(\theta_\text{sat}^\text{lb} - \theta) + (-mgl)(\sin(\theta_\text{sat}^\text{lb}) - \sin(\theta)) + \\ &\tau_{\text{human}}^{\text{max pf}} \cdot (\theta_\text{sat}^\text{lb} - \theta) + (-mgl)(\sin(\theta_\text{sat}^\text{ub}) - \sin(\theta_{\text{toeEq}})) \\
    & + mgl\cos(\theta_{\text{toeEq}})(\theta_\text{sat}^\text{ub} - \theta_{\text{toeEq}}).
\end{aligned}
\end{equation}
\section{Results}

\subsection{Stabilizable regions for age and sex adjusted models}
Table \ref{table:model_results} summarizes the results for the YF, YM, OM, OF, and wOF models with the GC and PD controller strategies. The changes in SR area for the forward and backward velocity regions are listed separately. In the forward angular velocity region, the GC and PD controllers slightly reduce the area of the SR in the young adult models but increase the total area in the OF and wOF model. The backward velocity region shows a different trend --- the exoskeleton assistance reduces SR area for all models. The additional damping provided by the PD controller does not significantly mitigate reductions in SR area relative to gravity compensation alone. The stabilizable regions for the YM and the wOF models are shown in Figure \ref{fig:SixPanel}a. 

As seen clearly in the figure, the addition of exoskeleton torque does not uniformly increase or decrease the area of the SR.
We see the following effects in the positive (posterior-to-anterior) velocity region:
\begin{itemize}
    \item The exo \textit{increases} the SR along the  entire backward fall boundary and along the forward fall boundary at low velocities.
    \item The exo \textit{decreases} the SR along the forward fall boundary at nondimensional velocities higher than approximately $0.2$.
\end{itemize}
At negative (anterior-to-posterior) velocities we see:
\begin{itemize}
    \item The exo \textit{increases} the SR along the forward fall boundary up to nondimensional velocities of approximately $-0.7$ and along the backward fall boundary at low velocities.
    \item the exo \textit{decreases} the SR along the backward fall boundary at velocities lower than approximately $-0.15$.
    \item the exo \textit{decreases} the  range of the contact constraint edge (the short edge in the lower right hand corner)
\end{itemize}

\begin{table} [h!] 
\centering
\caption{Percent change in total area of stabilizable region for sex- and age-adjusted models with exoskeleton assistance relative to no-exo baseline.}
\begin{tabular}{|c||c c | c c ||} 
 \hline
 & \multicolumn{2}{c|}{Gravity compensation} & \multicolumn{2}{c||}{Proportional-derivative}\\ [0.5ex] 
    & $v^+$ & $v^-$ & $v^+$ & $v^-$ \\
    \hline\hline
 YM & $ -0.4$ & $-16.9$ & $ -0.4$ & $-16.5$\\
 YF & $ -3.1$ & $-32.7$ & $ -2.7$ & $-32.4$\\ 
 OM & $ +3.7$ & $-15.4$ & $ +4.6$ & $-14.9$\\ 
 OF & $ +8.4$ & $-19.6$ & $ +9.6$ & $-18.9$\\
wOF & $+15.9$ & $-10.5$ & $+17.5$ & $-10.2$\\ [1ex] 
 \hline

\end{tabular}
\label{table:model_results}
\end{table} 
\subsection{Effect of reduced MT and MRTD}
\begin{figure}[h] 
    \centering
    \includegraphics[width=0.45\textwidth]{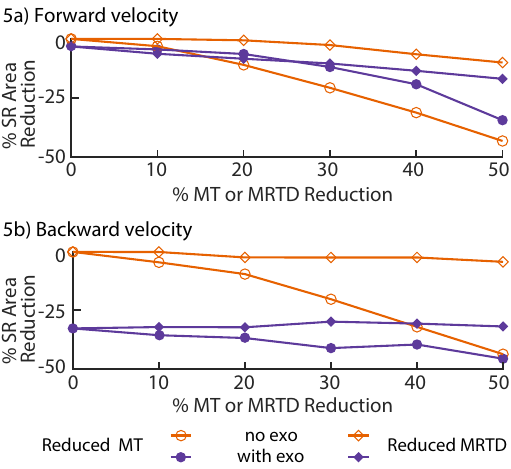}
    \caption{Trends in stabilizable region area when maximum torque (MT, circular markers) and maximum rate of torque development (MRTD, diamond markers) are independently reduced, with (filled/purple) and without the exoskeleton (open/orange). 5a) At forward velocities, the addition of the exoskeleton amplifies MRTD deficits, while it mitigates MT deficits, at least with respect to the \% SR area measure. 5b) At backward velocities, the exoskeleton tightens the foot-ground interaction constraints. This results in a large reduction in \% SR area, compared to the more subtle changes related to reduced MT or MRTD.}
    \label{fig:MTMRTDtrends}

\end{figure}

A graphical representation of the stabilizable region for YF, with and without exoskeleton assistance, when MT or MRTD are reduced by $30\%$ and $50\%$, is shown in Figures \ref{fig:SixPanel}b and \ref{fig:SixPanel}c. 
At positive CoM velocities, reducing maximum torque causes large reductions in SR area, but the addition of exoskeleton assistance mitigates the overall effect. For example, a $30\%$ MT reduction results in an SR area that is $20\%$ smaller than the baseline region. The SR area is only $10\%$ smaller when ankle exoskeleton assistance is included (Fig. \ref{fig:MTMRTDtrends}a). 
At negative velocities, the exoskeleton exacerbates MT deficits (Fig. \ref{fig:MTMRTDtrends}b).

When MRTD is reduced, the effect on the overall SR only becomes noticeable after large reductions ($\geq40\%$) at positive velocities. When an exoskeleton is added, however, the effect is amplified. Without the exoskeleton a $30\%$ reduction in MRTD causes a $2.6\%$ reduction in SR area. With an exoskeleton the SR area is reduced by $10.4\%$. At negative velocities, the change in total area is dominated by the reduction induced by simply adding the ankle exoskeleton.

\subsection{Work-energy bounds for the simpler model}
\begin{figure}[h] 
    \centering
    \includegraphics[width=0.45\textwidth]{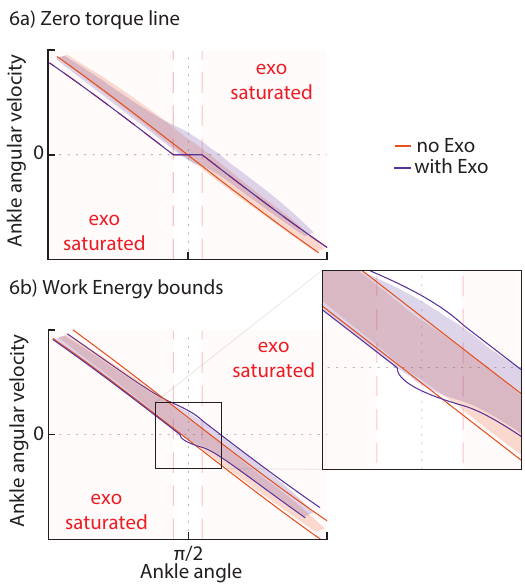} 
    \caption{Results of the work-energy (W-E) analysis. The zero torque line (ZTL) is shown in (6a) for the system without (orange line) and with the exoskeleton (purple line). Along these lines the human does not need to produce any torque to reach static equilibrium. Corresponding lines indicating the boundaries of the stabilizable region (SR), computed using W-E analysis, are shown in (6b). The SRs computed with our reachability-based method without (orange region) and with the exoskeleton (purple region) are shown for comparison. While the W-E analysis can capture some of the effects of ankle exoskeleton assistance, it cannot account for foot-ground interaction constraints or limited torque production capabilities and overapproximates the boundaries of the SR.}
    \label{fig:discussionFig}
\end{figure}
Near the upright posture and at low ankle angular velocities, the stabilizable regions for the exoskeleton and non-exoskeleton systems are well approximated by the bounds derived from the work-energy analysis (shown in Figure 6b for the wOF). At higher velocities and larger ankle angles, the work-energy bounds overapproximate the SR due to their inability to account for the foot-ground interaction constraints and the limited human MT and MRTD.

\section{Discussion}
In this study, we sought to characterize the effect of exoskeleton assistance at the ankle on feasible standing stability. We used simple, age- and sex-specific models, that reflect differences in human ankle torque production capabilities. The age related changes include reduced MT and MRTD. For each model, we computed the SR, consisting of all CoM positions and velocities from which it is feasible to recover to a quiet standing position without taking a step. We found that added ankle exoskeleton assistance with standard control strategies can increase feasible stability in low velocity nominal conditions for weaker models, but that it hinders stability for all models at higher CoM velocities.

The SRs computed without ankle exoskeleton assistance show that reduced MT has a far stronger effect on feasible stability than MRTD. Even a severe reduction in MRTD resulted in only a $10\%$ reduction in total SR area (Figure \ref{fig:MTMRTDtrends}). This aligns with results from the literature showing that the difference in plantar and dorsiflexor MRTD  between fallers and non-fallers is either weakly \cite{pijnappels2008} or not at all \cite{laroche2010} significant.
Interestingly, the stabilizable regions computed with ankle exoskeleton assistance suggest that the device may exacerbate some deficits. 
Reductions in MRTD lead to greater reductions in the total SR area when ankle exoskeleton assistance is included as compared to the baseline results without the exoskeleton 
(diamond shaped markers in Figure 5a and 5b). 
For reduced MT, the same detrimental effect is found only at negative (anterior-to-posterior) velocities; at positive (posterior-to-anterior) velocities, the ankle exoskeleton mitigates the reduction in SR area.

In full strength models, it is perhaps not surprising that the addition of the exoskeleton does not increase the area of the SR. These models have sufficiently high MT and MRTD bounds to maximize the size of the SR within the region where the foot-ground interaction constraints are not violated. The lack of change in total SR area at forward velocities (Table \ref{table:model_results}) aligns with the results from \cite{Emmens2018}, which showed no change in CoM kinematics as a result of moderate forward perturbations for young adults wearing ankle exoskeletons using the same controllers modeled here. 

At negative velocities, the SR of the full strength human-exoskeleton models is shrunk primarily along the short edge of the nominal SR (lower right quadrant of the subplots in Fig. \ref{fig:SixPanel}). This is due to the exoskeleton shifting the constraints on the total human torque (Equation \eqref{eq:torqueConstraintswithExo}). It must hold that 
$h_{\text{lb}}(x)-\tau_{\text{exo}} < x_3 < \tau^{\text{max df}}_{\text{human}}$ when the left hand side is  positive, but maximum (positive) dorsiflexion torque magnitude can be as low as $25\%$ of maximum (negative) plantar flexion torque, even in full strength models. This can result in constraint violations when $\tau_{\text{exo}}$ is large and negative.

In weaker models, the SR area measurement cannot fully describe the effects of an exoskeleton on standing balance as the exoskeleton does not uniformly alter the SR boundary (Fig. \ref{fig:SixPanel}). As noted in 
section \ref{Methods:Work-energy}, gravity sometimes provides an assistive torque in the no exoskeleton case. Therefore, if the exoskeleton controller is designed to cancel the gravitational torque, we might expect that exoskeleton ``assistance" does not uniformly improve stability. On the other hand, the exoskeleton increases the torque bounds of the human user. 
This is more important for weaker models, as stronger models can already produce enough torque to maximize feasible stability and are mainly limited by foot-ground interaction constraints.
In particular, an exoskeleton increases static stability in weaker models, resulting in a larger functional base of support.  A weak adult wearing the exoskeleton has the ability to do more mechanical work, but they must compensate at times for the gravitational torque being canceled out by the exoskeleton. 

The bounds computed using work-energy analysis approximate when this trade-off, between the exoskeleton providing mechanical work vs. requiring users to do additional mechanical work, 
occurs. Thus, these bounds indicate regions in the phase plane where exoskeleton assistance drives the reduction in feasible stability (Figure \eqref{fig:discussionFig}). Work-energy bounds are, however, far less precise than the stabilizable regions, because the work-energy analysis assumes conservation of energy, ignores ground contact and input constraints, and cannot generalize to systems with dissipation (such as exoskeleton controllers with damping). Together, the higher and lower dimensional models provide a unique way to understand the mechanisms through which ankle exoskeletons may affect standing balance. As we find a benefit to having an exoskeleton in low velocity conditions, but a corresponding cost at high velocities, a major implication of our work is that while designing exoskeleton controllers, we must consider what nominal conditions and extreme scenarios older users might experience and then test these devices accordingly. There may be unintended consequences from an assistive device that aids stability in most cases if it increases the risk of falls in even a small percentage of scenarios a user may experience.

A limitation of this study is the simplicity of the models. Our results generate bounds on ankle strategies for balance, but they do not allow the use of hip or knee strategies that could enhance the SR. In that sense, our results are likely an under-approximation of the true SR. We model both the human and exoskeleton torque simply, not accounting for nonlinear muscle dynamics, physiological actuation delays, device delays, and the physical interface between the human body and the device. The interaction between physiological delay and exoskeleton actuation delay, in particular, likely affects ankle exoskeleton-assisted standing balance \cite{Beck2023,Sharafi2023}, so we plan to include delay in future modeling work.

An interesting result of the work-energy analysis relates to the shifts in the ZTL when the exoskeleton is added. 
Examining the ZTL with the exoskeleton (Fig. \ref{fig:discussionFig}a, purple lines), we can see that in the forward velocity half of the phase plane the line is shifted to the left of the no-exoskeleton ZTL (Fig. \ref{fig:discussionFig}a, orange line). Thus, a plantar flexion braking torque is always required to prevent forward falls. The addition of the exoskeleton effectively eliminates the possibility of a backward fall occurring, as long as there is an initial positive velocity and the ground contact constraints are satisfied. In the negative velocity half of the plane, the ZTL is shifted to the right, thereby increasing the area where a dorsiflexion braking torque is needed to prevent a backward fall.
These shifts in the ZTL mean that the exoskeleton qualitatively changes the optimal human response required to stabilize to quiet standing in large regions of the state space. This may have unforeseen implications on the ability of users to adapt to exoskeleton assistance, preventing them from responding with optimal control strategies. We plan to explore this idea further by modeling suboptimal adaptation to these devices, which could help guide future empirical testing of device controllers.

It is also important to analyze more sophisticated controller strategies. The small reduction in area in the full-strength YM model results from having a large enough range of positive and negative MT and large enough MRTD such that undesireable exoskeleton behavior can be compensated for quickly. An exoskeleton that compensates for missing MT and MRTD  by directly amplifying ankle torque could enable older users to uniformly enlarge their SR. On the other hand, such a strategy could unintentionally amplify suboptimal human control strategies. This approach is often implemented as proportional myoelectric control \cite{ferris2009}. To enable analysis of such myoelectric controllers, we will include more detailed muscle actuation models in our future work. 

Lastly, we hope to develop model-based experimental protocols. To fully understand the effect of ankle exoskeletons on stability, the stabilizable regions suggest that it is insufficient to apply relatively moderate forward perturbations that are common in experiments \cite{Emmens2018, Beck2023}. Our regions can serve as a guide in the design of perturbation experiments of older adults wearing exoskeletons. Given the torque production capabilities of a participant, the perturbation magnitude and direction required to sufficiently challenge stability can be calculated. This will enable thorough, model-driven experimental evaluations of ankle-exoskeleton assisted standing balance.

\section*{Acknowledgment}

Thanks to Anat Raman for the many fruitful discussions. 

\bibliographystyle{ieeetr}
\bibliography{refs}

\begin{thebibliography}{10}

\bibitem{Poggensee2021}
K.~L. Poggensee and S.~H. Collins, ``How adaptation, training, and customization contribute to benefits from exoskeleton assistance,'' {\em Science Robotics}, vol.~6, p.~eabf1078, Sep 2021.

\bibitem{Lakmaa2024}
A.~Lakmazaheri, S.~Song, B.~B. Vuong, B.~Biskner, D.~M. Kado, and S.~H. Collins, ``Optimizing exoskeleton assistance to improve walking speed and energy economy for older adults,'' {\em Journal of NeuroEngineering and Rehabilitation}, vol.~21, p.~1, Jan 2024.

\bibitem{Raitor2024}
M.~Raitor, S.~W. Ruggles, S.~L. Delp, C.~K. Liu, and S.~H. Collins, ``Lower-limb exoskeletons appeal to both clinicians and older adults, especially for fall prevention and joint pain reduction,'' {\em IEEE Transactions on Neural Systems and Rehabilitation Engineering}, vol.~32, pp.~1577--1585, Mar 2024.

\bibitem{horak1986}
F.~B. Horak and L.~M. Nashner, ``Central programming of postural movements: adaptation to altered support-surface configurations,'' {\em Journal of neurophysiology}, vol.~55, pp.~1369--1381, Jun 1986.

\bibitem{Hasson2014}
C.~J. Hasson, R.~E. van Emmerik, and G.~E. Caldwell, ``Balance decrements are associated with age-related muscle property changes,'' {\em Journal of Applied Biomechanics}, vol.~30, pp.~555--562, Aug 2014.

\bibitem{Cattagni2014}
T.~Cattagni, G.~Scaglioni, D.~Laroche, J.~Van~Hoecke, V.~Gremeaux, and A.~Martin, ``Ankle muscle strength discriminates fallers from non-fallers,'' {\em Frontiers in Aging Neuroscience}, vol.~6, p.~336, Dec 2014.

\bibitem{CATTAGNI2016}
T.~Cattagni, G.~Scaglioni, D.~Laroche, V.~Gremeaux, and A.~Martin, ``The involvement of ankle muscles in maintaining balance in the upright posture is higher in elderly fallers,'' {\em Experimental Gerontology}, vol.~77, pp.~38--45, May 2016.

\bibitem{Thelen1996}
D.~G. Thelen, A.~B. Schultz, N.~B. Alexander, and J.~A. Ashton-Miller, ``{Effects of Age on Rapid Ankle Torque Development},'' {\em The Journals of Gerontology: Series A}, vol.~51A, pp.~M226--M232, Sep 1996.

\bibitem{bento2010}
P.~C.~B. Bento, G.~Pereira, C.~Ugrinowitsch, and A.~L.~F. Rodacki, ``Peak torque and rate of torque development in elderly with and without fall history,'' {\em Clinical biomechanics}, vol.~25, pp.~450--454, Jun 2010.

\bibitem{kamo2019}
T.~Kamo, R.~Asahi, M.~Azami, H.~Ogihara, T.~Ikeda, K.~Suzuki, and Y.~Nishida, ``Rate of torque development and the risk of falls among community dwelling older adults in japan,'' {\em Gait \& Posture}, vol.~72, pp.~28--33, Jul 2019.

\bibitem{ema2016}
R.~Ema, M.~Saito, S.~Ohki, H.~Takayama, Y.~Yamada, and R.~Akagi, ``Association between rapid force production by the plantar flexors and balance performance in elderly men and women,'' {\em Age}, vol.~38, pp.~475--483, Dec 2016.

\bibitem{laroche2010}
D.~P. LaRoche, K.~A. Cremin, B.~Greenleaf, and R.~V. Croce, ``Rapid torque development in older female fallers and nonfallers: a comparison across lower-extremity muscles,'' {\em Journal of electromyography and kinesiology}, vol.~20, pp.~482--488, Jun 2010.

\bibitem{pijnappels2008}
M.~Pijnappels, J.~Van~der Burg, N.~D. Reeves, and J.~H. van Die{\"e}n, ``Identification of elderly fallers by muscle strength measures,'' {\em European journal of applied physiology}, vol.~102, pp.~585--592, Mar 2008.

\bibitem{Beck2023}
O.~N. Beck, M.~K. Shepherd, R.~Rastogi, G.~Martino, L.~H. Ting, and G.~S. Sawicki, ``Exoskeletons need to react faster than physiological responses to improve standing balance,'' {\em Science Robotics}, vol.~8, p.~eadf1080, Feb 2023.

\bibitem{Sharafi2023}
S.~Sharafi and T.~K. Uchida, ``Stability of human balance during quiet stance with physiological and exoskeleton time delays,'' {\em IEEE Robotics and Automation Letters}, vol.~8, pp.~6211--6218, Aug 2023.

\bibitem{RAO2006}
N.~Rao and A.~S. Aruin, ``Automatic postural responses in individuals with peripheral neuropathy and ankle--foot orthoses,'' {\em Diabetes Research and Clinical Practice}, vol.~74, pp.~48--56, Oct 2006.

\bibitem{Canete2023}
S.~Canete, E.~B. Wilson, W.~G. Wright, and D.~A. Jacobs, ``The effects of exoskeleton assistance at the ankle on sensory integration during standing balance,'' {\em IEEE Transactions on Neural Systems and Rehabilitation Engineering}, vol.~31, pp.~4428--4438, Nov 2023.

\bibitem{fang2023}
Y.~Fang and Z.~F. Lerner, ``How adaptive ankle exoskeleton assistance affects stability during perturbed and unperturbed walking in the elderly,'' {\em Annals of Biomedical Engineering}, vol.~51, no.~11, pp.~2606--2616, 2023.

\bibitem{Son2010}
J.~Son and J.~Ashton-Miller, ``Do ankle orthoses improve ankle proprioceptive thresholds or unipedal balance in older persons with peripheral neuropathy?,'' {\em American Journal of Physical Medicine and Rehabilitation}, vol.~89, pp.~369--375, May 2010.

\bibitem{Emmens2018}
A.~R. Emmens, E.~H.~F. van Asseldonk, and H.~van~der Kooij, ``Effects of a powered ankle-foot orthosis on perturbed standing balance,'' {\em Journal of NeuroEngineering and Rehabilitation}, vol.~15, p.~50, Dec 2018.

\bibitem{PAIPATTON1997}
Y.~C. Pai and J.~Patton, ``Center of mass velocity position predictions for balance control,'' {\em Journal of Biomechanics}, vol.~30, pp.~347--354, Apr 1997.

\bibitem{yang2009}
F.~Yang, D.~Espy, and Y.~C. Pai, ``Feasible stability region in the frontal plane during human gait,'' {\em Annals of biomedical engineering}, vol.~37, pp.~2606--2614, Dec 2009.

\bibitem{bahari2021}
H.~Bahari, J.~Forero, J.~C. Hall, J.~S. Hebert, A.~H. Vette, and H.~Rouhani, ``Use of the extended feasible stability region for assessing stability of perturbed walking,'' {\em Scientific Reports}, vol.~11, p.~1026, Jan 2021.

\bibitem{HOF20051}
A.~Hof, M.~Gazendam, and W.~Sinke, ``The condition for dynamic stability,'' {\em Journal of Biomechanics}, vol.~38, pp.~1--8, Jan 2005.

\bibitem{Raz2023}
D.~Raz, L.~Yang, B.~R. Umberger, and N.~Ozay, ``Determining the domain of stable human sit-to-stand motions via controlled invariant sets and backward reachability,'' in {\em 2023 European Control Conference (ECC)}, pp.~1--7, Jun 2023.

\bibitem{mummolo2017}
C.~Mummolo, L.~Mangialardi, and J.~H. Kim, ``Numerical estimation of balanced and falling states for constrained legged systems,'' {\em Journal of Nonlinear Science}, vol.~27, pp.~1291--1323, Aug 2017.

\bibitem{orsolino2020}
R.~Orsolino, M.~Focchi, S.~Caron, G.~Raiola, V.~Barasuol, D.~G. Caldwell, and C.~Semini, ``Feasible region: An actuation-aware extension of the support region,'' {\em IEEE Transactions on Robotics}, vol.~36, pp.~1239--1255, Jun 2020.

\bibitem{Inkol2022}
K.~A. Inkol and J.~McPhee, ``Using dynamic simulations to estimate the feasible stability region of feet-in-place balance recovery for lower-limb exoskeleton users,'' in {\em 2022 9th IEEE RAS/EMBS International Conference for Biomedical Robotics and Biomechatronics (BioRob)}, pp.~1--6, Aug 2022.

\bibitem{watson2021}
F.~Watson, P.~C. Fino, M.~Thornton, C.~Heracleous, R.~Loureiro, and J.~J. Leong, ``Use of the margin of stability to quantify stability in pathologic gait--a qualitative systematic review,'' {\em BMC musculoskeletal disorders}, vol.~22, pp.~1--29, 2021.

\bibitem{vukobratovic2004}
M.~Vukobratovi{\'c} and B.~Borovac, ``Zero-moment point—thirty five years of its life,'' {\em International journal of humanoid robotics}, vol.~1, pp.~157--173, Mar 2004.

\bibitem{Mitchell2005}
I.~Mitchell, A.~Bayen, and C.~Tomlin, ``A time-dependent hamilton-jacobi formulation of reachable sets for continuous dynamic games,'' {\em IEEE Transactions on Automatic Control}, vol.~50, pp.~947--957, Jul 2005.

\bibitem{Mo2018}
M.~Chen and C.~J. Tomlin, ``Hamilton--jacobi reachability: Some recent theoretical advances and applications in unmanned airspace management,'' {\em Annual Review of Control, Robotics, and Autonomous Systems}, vol.~1, no.~1, pp.~333--358, 2018.

\bibitem{ERA2006}
P.~Era, P.~Sainio, S.~Koskinen, P.~Haavisto, M.~Vaara, and A.~Aromaa, ``{Postural Balance in a Random Sample of 7,979 Subjects Aged 30 Years and Over},'' {\em Gerontology}, vol.~52, pp.~204--213, Jul 2006.

\bibitem{sontag1983}
E.~D. Sontag, ``A lyapunov-like characterization of asymptotic controllability,'' {\em SIAM journal on control and optimization}, vol.~21, pp.~462--471, May 1983.

\bibitem{Endo2002}
M.~Endo, J.~A. Ashton-Miller, and N.~B. Alexander, ``{Effects of Age and Gender on Toe Flexor Muscle Strength},'' {\em The Journals of Gerontology: Series A}, vol.~57, pp.~M392--M397, Jun 2002.

\bibitem{ferris2009}
D.~P. Ferris and C.~L. Lewis, ``Robotic lower limb exoskeletons using proportional myoelectric control,'' in {\em 2009 Annual international conference of the Ieee engineering in medicine and biology society}, pp.~2119--2124, IEEE, 2009.

\end{thebibliography}

\appendices
\section{Derivation of Equations of Motion and Constraints} \label{Apdx:EoMandConst}
Here we derive the equations of motion and state constraints presented in Section III.
\subsection{Equations of Motion}
\begin{figure}[h] 
    \centering
    \includegraphics[width=0.4\textwidth]{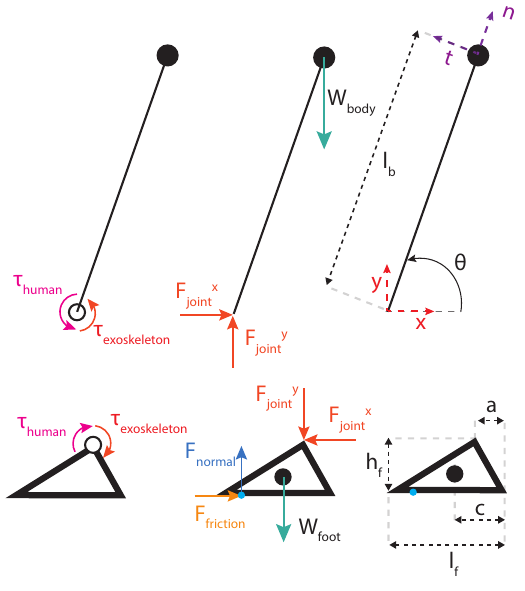}
    \caption{Free body diagram of human-exoskeleton system showing the forces and moments acting on the body and foot segment }
    \label{fig:FBD_LS}
\end{figure}

A free body diagram of the system is shown in Figure \ref{fig:FBD_LS}. To derive the constraints that prevent the foot from slipping or tipping, we must first understand what torques and forces are acting on the foot.

In the following, we will denote $\cos{(\theta)}$ as $c_\theta$ and $\sin{(\theta)}$ as $s_\theta$ for particularly unwieldy equations. Let $\tau = \tau_{\text{human}} + \tau_{\text{exoskeleton}}$ be the total joint torque at the ankle. Summing the torques for the body segment at the ankle, we have
\begin{equation}
    \tau = ml_b^2\Ddot{\theta} + mgl_b\cos{\theta},
    \label{torqueDynamics}
\end{equation}
where $ml_b^2$ is the moment of inertia of the body segment.
The rotational motion of the body segment can be decomposed into a normal acceleration along the leg
\begin{equation}
    a^n = -l_b\Dot{\theta}^2
    \label{Fc}
\end{equation} 
and a tangential acceleration
\begin{equation}
    a^t = l_b\Ddot{\theta}.
    \label{Ft}
\end{equation} 
Rotating these accelerations to the x-y coordinate system, we have - 
\begin{equation}
    a^x = a^n c_\theta - a^t s_\theta = -l_b(\Dot{\theta}^2c_\theta + \Ddot{\theta}s_\theta)
\end{equation}

\begin{equation}
    a^y = a^n s_\theta + a^t c_\theta = l_b( - \Dot{\theta}^2 s_\theta + \Ddot{\theta} c_\theta)
\end{equation}

These accelerations are caused by a combination of the weight of the body and the contact forces at the ankle joint. Thus we can write the force balance at the ankle as:
\begin{equation}
    m a^x = F_\text{joint}^x = -m l_b(\Dot{\theta}^2 c_\theta + \Ddot{\theta} s_\theta)
\end{equation}

\begin{equation}
    m a^y = F_\text{joint}^y - W_\text{body} = F_\text{joint}^y - mg = m l_b( - \Dot{\theta}^2 s_\theta + \Ddot{\theta} c_\theta)
\end{equation}
Similarly, for the foot segment we can write the force balance as:
\begin{equation}
    m_f a_f^x = F_\text{friction} - F_\text{joint}^x
\end{equation}
\begin{equation}
    m_f a_f^y = F_\text{normal} - F_\text{joint}^y - W_\text{foot} =  F_\text{normal} - F_\text{joint}^y - m_fg
\end{equation}
where $m_f$ is the mass of the foot. When the foot is not accelerating we get:
\begin{equation}
    F_\text{normal} = F_\text{joint}^y + m_fg
\end{equation}
This gives us an equation for $F_\text{normal}$:
\begin{equation} \label{eq:F_normal}
    F_\text{normal} = ml_b(-\Dot{\theta}^2 s_\theta +\Ddot{\theta}c_\theta) + mg + m_fg
\end{equation}
Similarly, we can derive an expression for the ground reaction/friction force in the $x$ direction via force balancing:
\begin{equation}
    F_\text{friction} = F_\text{joint}^x = -ml_b(\Dot{\theta}^2 c_\theta +\Ddot{\theta} s_\theta)
\end{equation}

Finally, the location of the center of pressure (CoP) as a function of these known quantities can be found by balancing torques about the ankle:
\begin{equation}
   ((l_f - a) -\text{CoP})\cdot F_\text{normal} - cm_fg + \tau - h_fF_\text{friction} = 0
\end{equation}
which in turn gives us
\begin{equation}
    \text{CoP} = (l_f - a) - \frac{cm_fg - \tau + h_f F_\text{friction}}{F_\text{normal}}
    \label{CopEq}
\end{equation}
These state-dependent equations for the CoP and the forces acting on the foot are useful for expressing foot-ground reaction constraints in terms of state, as will be shown in the next section.
\subsection{State Constraints}
Constraints (C3) and (C5), represent the ground reaction force and friction force constraints that must be satisfied for the foot to maintain contact with the ground. To incorporate them into our computational framework, they must be converted into constraints on the system state, which includes the applied human torque at the ankle. We first derive constraints on $\tau$, which we later convert to constraints on the human and exoskeleton torque in the body of the paper, by solving \eqref{torqueDynamics} for the acceleration:
\begin{equation}
    \Ddot{\theta} = \frac{1}{ml_b^2}\tau - \frac{g}{l_b}c_\theta
    \label{accelDynamics}
\end{equation}
For Constraint \eqref{GRFconstraint}, substituting for $\Ddot{\theta}$ into \eqref{eq:F_normal} gives us
\begin{equation}
 ml(-\Dot{\theta}^2s_\theta +(\frac{1}{ml^2}\tau - \frac{g}{l}c_\theta)c_\theta) + mg + m_fg\geq 0.
\end{equation}
Solving for $\tau$, after some algebra we see that 

\begin{equation}
   \tau  \geq \frac{ -l(m + m_f)g + ml^2\Dot{\theta}^2s_\theta}{c_\theta} +  mglc_\theta
\end{equation}
if $c_\theta > 0$, with the direction of the inequality flipped if $c_\theta < 0$.

Similarly, plugging \eqref{accelDynamics} into constraint (C5) results in four inequalities constraining $\tau$. Letting $\rho_1 = \mu c_\theta+ s_\theta$ and $\rho_2 = \mu c_\theta + s_\theta$, when $\rho_1 < 0$
\begin{equation}
    \tau < mglc_\theta + \frac{(\mu s_\theta - c_\theta)ml^2\Dot{\theta}^2 - \mu lg(m_f+m)}{\rho_1}
\end{equation}
with the direction of the inequality flipped when $\rho_1 > 0$. When $\rho_2 < 0$,
\begin{equation}
        \tau < mglc_\theta + \frac{(\mu s_\theta + c_\theta)ml^2\Dot{\theta}^2 - \mu lg(m_f+m)}{\rho_2}
\end{equation}

Lastly, constraints on $\tau$ are derived from \eqref{CoPconstraint} by first incorporating the parameters of the foot:
\begin{equation}
    0 \leq  \text{CoP} \leq l_f
\end{equation}
Plugging in our expression for the CoP, equation \eqref{CopEq}, gives us two corresponding inequalities:
\begin{equation}
      \tau  <  aF_\text{normal} + h_f F_\text{friction} + cm_fg
\end{equation}
and
\begin{equation}
    \tau > -(l_f - a )F_\text{normal} + cm_fg + h_f F_\text{friction}
\end{equation}
We now have state-dependent constraints on the total joint torque $\tau$, which can be further broken down into constraints on the human and exoskeleton torque. This representation of the constraints is thus compatible with the backwards reachability analysis described in the Methods section.
\section{Target Set Computation} \label{apdx:CLF}
Our approach for computing controlled invariant target sets leverages Control Lyapunov Functions (CLFs) to generate invariant ellipsoids. We first construct the ellipsoids, and then scale them such that desired linear constraints on state and input are satisfied.
\subsection{Computing the CLF} 
To synthesize a CLF of the form $E(x) = x^TPx$  with corresponding state feedback controller $u=Kx$, we solve the following optimization problem, where both constraints come from Definition 4 in the main text:
\begin{equation}
    \begin{aligned}
    \min_{P,K}\quad & \textrm{ Tr}(P) \\
    \textrm{s.t.} \quad & P > 0 \\
    \quad & (A + BK)^TP + P(A + BK) < 0 \\
\end{aligned}
\end{equation}
This problem is nonlinear in $K$ and $P$, but it can be converted to a convex SDP. First, we multiply the second constraint by $P^{-1}$ on both sides, so that it becomes
\begin{equation}
    P^{-1}((A + BK)^T + (A + BK))P^{-1} < 0.
\end{equation}
Letting $Q = P^{-1}$ and $Y = KQ$, we can write an SDP with linear constraints
\begin{equation}
\begin{aligned}
    \min_{Q,Y}\quad & \textrm{ Tr}(Q) \\
    \textrm{s.t.} \quad & Y^TB^T+QA^T + BY < 0.
\end{aligned}
\end{equation}
The solution gives CLF $E(x) = x^TPx$, where $P = Q^{-1}$, and a controller $u = Kx$, where $K = YP$.
\subsection{Computing maximally scaled ellipsoids}
Now that we have constructed a ellipsoid CLF, the target set will be a union of sets of the form $\{x \mid (x-x_\text{eq})^TP(x-x_\text{eq}) \leq c\} $ for equilibrium point $x_{\text{eq}}$ and some $c \in \mathbb{R}$. We must scale each ellipsoid by selecting $c$ such that state and input constraints are satisfied. 

To simplify this problem, we first consider only linear constraints. The controller $u = Kx$ should not violate constraint (C2) and the constraint on the human torque production (C1) should also be enforced. Furthermore, the target set should not contain CoM positions anterior or posterior to the foot. As the target set should represent quiet standing, we also add an additional constraint limiting the velocity within the target by an amount close to experimentally observed postural sway (see [32] for example). These last two conditions become two additional linear constraints:
\begin{align}
\theta_{\text{heel}} &< x_1 < \theta_{\text{toe}} \tag{C6} \\
   -v_{\text{sway}} & < x_2 <  v_{\text{sway}} \tag{C7}
    \end{align}
where $v_{\text{sway}}$ represents the maximum allowable postural sway within the target set.

For consistency and computational ease, we will choose the largest $c$ such that the ellipsoid $P$ is maximally contained within the linear constraints. Finding this value is just a simplified version of the convex problem of finding a maximum volume inscribed ellipsoid within a polyhedron (see for example Chapter 8 of Boyd and Vandenberghe's \textit{Convex Optimization}) . It has a closed form solution which we will derive.

Let $(x-x_\text{eq})^TP(x-x_\text{eq})$ be the CLF corresponding to a given equilibrium point. Let $A_\text{con}x = b_\text{con}$ represent a set of $n$ linear constraints. We wish to find the largest $c$ such that the ellipsoid $\mathcal{E} = \{x \vert (x-x_\text{eq})^TP(x-x_\text{eq}) \leq c\}$ is fully contained within the constraints. Denote $a_i$, $b_i$ the $i$th row of $A_\text{con}$ and entry of $b_\text{con}$, respectively. In the case of a single constraint hyperplane $a_i^Tx \leq b_i$, the largest possible $c$ should produce an ellipsoid that intersects the constraint at one point. This can be cast as a simple convex optimization problem:
\begin{align}
    \min_x \quad  &(x-x_\text{eq})^TP(x-x_\text{eq}) \\
    \textrm{s.t.} \quad 
    &a_i^Tx = b_i \nonumber
\end{align}
By the method of Lagrange multipliers, we have that the Lagrangian $L(x) = (x-x_\text{eq})^TP(x-x_\text{eq}) + \lambda (a_i^Tx - b_i)$. Taking the partial derivatives, we solve
\begin{equation}
\begin{aligned}
    2P(x-x_\text{eq}) + \lambda a_i^T &= 0,\\ a_i^Tx - b_i &= 0 \\
\end{aligned}
\end{equation}
for $\lambda, x$. After some algebra this gives an expression for $c$, which in turn should be minimized over all $n$ constraints:

\begin{equation}
    c = \min_i \, \frac{(b_i - a_i^Tx_\text{eq})^2}{a_i^TP^{-1}a_i}
\end{equation}

For compatibility with the BRS computation tool, we enclose the resulting ellipsoid in a tight axis-aligned bounding box. We check that the vertices are also contained within the constraint polytope. If they are not, we perform a linear search to further scale $c$ such that the constraints are satisfied for the minimum circumscribed axis-aligned bounding box. 
To account for the nonlinear foot-ground interaction constraints, we check that the box corners do not violate \eqref{GRFconstraint}-\eqref{frictionConstraint}. 
Although this does not provide a provable guarantee of constraint satisfaction, in practice it is a sufficient check due to the relatively small size of the boxes.
The target set is then the union of the axis-aligned boxes centered around the  ankle and toe equilibria. 
\section{Computational Considerations}

The code used to produce the results in our tables and figures is hosted at \url{https://github.com/Raquelometer/AnkleExoBRS}. To improve computational performance, the states of system \eqref{eq:Lifted_IP_dynamics} were scaled to make them dimensionless. Angular velocities $\Dot{\theta}$ were scaled by $\omega = \sqrt{l/g}$, the natural frequency of the pendulum. The human torque state $\tau_h$ was non-dimensionalized by multiplying by $\frac{\omega^2}{ml^2}$.

The backward reachable set was computed in the dimensionless state space over a $225 \times 225 \times 225$ grid. This was the coarsest grid which produced sufficiently clean results. We found that results computed over coarser grids were too noisy.

Due to computational scaling issues we compute the backward reachable sets in the space of positive (posterior to anterior) and negative (anterior to posterior) velocities separately.
Segmenting the state space in this manner means that the backward reachable sets may not include initial conditions whose resulting trajectories oscillate between (large) positive and negative velocities while satisfying the constraints. 
To mitigate this issue within reasonable bounds for human balance, we include a buffer within our velocity range such that we compute over $[v^-_{\text{buffer}}, v^+_{\text{lb}}]$. We select the size of this buffer to be $\SI{0.5}{\radian\per\second}$, which is the largest computationally feasible bound. 
Our backwards reachable sets therefore exclude true failure states --- from which it is impossible to reach the target set, and high oscillation states --- from which it is only possible to reach the target set with oscillations of a magnitude higher than $v_{\text{buffer}}$. We note that in practice the size of the velocity buffer does not affect the resulting BRS and it is likely that highly oscillatory solutions do not satisfy constraints. 

Because we compute the BRS in two separate forward and backward velocity segments, the target set formulation must encompass all the portions of the foot that we wish to reach with mostly positive and mostly negative velocity only.
 We therefore make the following observation: If an equilibrium point near the toe can be reached with a positive velocity, then any point posterior to that can be eventually reached with a low enough negative velocity, i.e. by leaning back slowly. 
 This implies that it is sufficient to select one equilibrium point at the toe, rather than having to represent the entire range of static equilibria. 
 A symmetric argument can be made for an equilibrium point at the heel or ankle for points that can be reached with a negative velocity. Therefore the ankle and toe equilibria selected in IV-B are appropriate.
 
\end{document}